\documentclass[12pt,reqno]{article}
\usepackage[a4paper]{geometry}
\usepackage{amsmath,amssymb,amsfonts}
\usepackage{graphicx}
\usepackage[margin=10pt,font=small]{caption}

\usepackage{slashed}

\relax

\def\be{\begin{equation}}
\def\ee{\end{equation}}
\def\bea{\begin{eqnarray}}
\def\eea{\end{eqnarray}}

\newcommand\fverb{\setbox\pippobox=\hbox\bgroup\verb}
\newcommand\fverbdo{\egroup\medskip\noindent%
                        \fbox{\unhbox\pippobox}\ }
\newcommand\fverbit{\egroup\item[\fbox{\unhbox\pippobox}]}

\newcommand{\la}{\lambda}

\newcommand{\bear}{\begin{eqnarray}}

\newcommand{\eear}{\end{eqnarray}}

\newbox\pippobox

\def\lab{\label}
\def\6{\partial}

\def\a{\alpha}

\def\half{\frac12}

\def\le{\left}
\def\ri{\right}
\def\cO{{\cal O}}

\def\C0{{\bf C_0}}
\def\Y0{{\bf Y_0}}
\def\G0{{\bf G_0}}

\def\m{\mu}

\def\sq
\def\a{\alpha}

\def\l{\lambda}

\def\tr{{\rm Tr}}

\def\eps{\epsilon}

\def\la{\langle}
\def\ra{\rangle}

\def\bz{\begin{itemize}}
\def\ez{\end{itemize}}
\def\bn{\begin{enumerate}}
\def\en{\end{enumerate}}

\def\ol{\overline}
\def\ben{\begin{enumerate}}
\def\een{\end{enumerate}}

\def\a{\alpha}

\def\6{\partial}
\def\lab{\label}
\def\half{\frac12}
\def\le{\left}
\def\ri{\right}

\def\be{\begin{equation}}
\def\ee{\end{equation}}
\def\bea{\begin{eqnarray}}
\def\eea{\end{eqnarray}}
\def\bz{\begin{itemize}}
\def\ez{\end{itemize}}
\def\la{\langle}
\def\ra{\rangle}

\def\cO{{\cal O}}

\allowdisplaybreaks[2]
\numberwithin{equation}{section}

\newcommand{\eq}[1]{\begin{equation}
                     \begin{split} #1 \end{split}
                     \end{equation}}

\newcommand{\ov}[1]{\overline{#1}}
\newcommand{\ul}[1]{\underline{#1}}

\begin{document}
\begin{flushright} \small
 ITP--UU--11/47 \\ SPIN--11/37 \\ CERN-PH-TH/2011-321
\end{flushright}

\vskip1cm

\begin{center}
 {\large\bfseries Holography and ARPES sum-rules}\\

 \vskip0.8cm

Umut G\"ursoy\textsuperscript{\dag}, Erik Plauschinn\textsuperscript{*},
Henk Stoof\hspace{1.5pt}\textsuperscript{*}, Stefan Vandoren\textsuperscript{*} \\

\vskip0.8cm

{\small\slshape
\textsuperscript{\dag} Theory Group, Physics Department, CERN \\ CH-1211 Geneva 23, Switzerland\\
 \medskip
 \textsuperscript{*} Institute for Theoretical Physics and Spinoza Institute,
 Utrecht University \\ 3508 TD Utrecht, The Netherlands \\

\vskip0.8cm

{\upshape\ttfamily Umut.Gursoy@cern.ch; E.Plauschinn, H.T.C.Stoof, S.J.G.Vandoren@uu.nl}\\[3mm]}
\end{center}

\vspace{10mm}
\hrule
\vspace{5mm}
\centerline{\bfseries Abstract}
\vskip0.2cm
\noindent
We study correlation functions of elementary fermions in strongly interacting field theories using the AdS/CFT correspondence. This correspondence generically associates bulk fields to composite operators in field theory. We modify the holographic prescription in order to obtain correlators that correspond to fermonic single-particle excitations by introducing a dynamical fermionic source localized on a UV brane in a holographic background. 
 We work out the conditions when these correlators obey the zeroth frequency sum-rule satisfied by angle-resolved photo-emission spectroscopy (ARPES) and are thus directly relevant to the AdS/CMT correspondence. To illustrate our techniques, we study field theories at zero chemical potential with an arbitrary dynamical exponent $z$, i.e., the Lifshitz invariant conformal field theories, including the usual relativistic case $z=1$.

\medskip

\bigskip
\hrule\bigskip
\newpage


\section{Introduction}\lab{sec:intro}

The AdS/CMT program, see \cite{review} and references therein, is a useful tool for the description of strongly interacting condensed-matter systems that  grew out of the AdS/CFT correspondence \cite{AdSCFT} recently. 
In its first instance it entails a duality between a gravity theory in a
 $(d+1)$-dimensional anti-de Sitter (AdS) bulk space-time and a
strongly interacting conformal quantum field theory (CFT) living on the $d$-dimensional
boundary of the AdS space. 
Perturbations away from the exact quantum critical point
described by the conformal field theory
can be realized, for instance,  by considering a black brane in the
anti-de Sitter space-time
leading to a nonzero temperature for the theory on the boundary.
Furthermore, a black brane charged with respect to 
an electro-magnetic gauge field corresponds  to adding a chemical potential
to the CFT. Studies of fermionic correlation functions in these systems lead to interesting Fermi and non-Fermi-like behavior \cite{SSLee,MIT1,MIT2,Leiden}. 

\begin{figure}[h]
\centering
\vskip10pt
\includegraphics[width=0.6\textwidth]{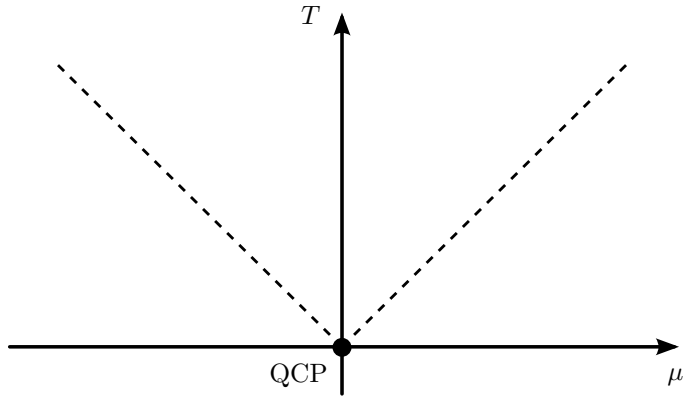}
\begin{picture}(0,0)
\put(-145,140){\footnotesize $T$}
\put(-8,6){\footnotesize $\mu$}
\put(-157,5){\footnotesize QCP}
\end{picture}
\caption{The phase diagram of a particle-hole symmetric
semi-metal. The temperature of the semi-metal is denoted by $T$ and
the chemical potential, which determines the nature and amount of
doping of the semi-metal, by $\mu$. The dashed lines denote the
smooth crossover between the classical and quantum behavior of the
normal phase of the semi-metal. The latter is governed by the
quantum critical point (QCP) at zero temperature and zero chemical
potential.}
\end{figure}

To understand that the AdS/CFT correspondence may also be useful
for con\-densed-matter physics, let us revisit the phase diagram of
a particle-hole symmetric semi-metal. It is shown in figure 1, in a manner convenient for this
discussion. Plotting the phase
diagram in this way clearly shows  that the behavior of the
semi-metal is governed by a quantum critical point at zero
temperature and zero chemical potential. 
Within the context of the AdS/CFT correspondence, the conformal field theory associated with this quantum critical point
has a particular important feature, namely that under a scale
transformation the time direction scales equally fast as the
spatial directions. This is due to the relativistic invariant
nature of the anti-de Sitter background, and from the point of view
of condensed-matter physics it corresponds to a quantum critical
point that has a dynamical exponent $z=1$. Interestingly, graphene
presents a concrete realization of this situation.
However, in condensed-matter physics one more often deals with the
situation that under a scale transformation the time direction
scales twice as fast as the spatial directions, i.e., the
dynamical exponent is $z=2$. Examples are double-layer graphene
or gapless semi-conductors. The latter are usually not
exactly particle-hole symmetric, but in a first approximation this
will not change the physics qualitatively. A system with $z=2$
that breaks particle-hole symmetry in a more fundamental manner is
given by an atomic Fermi mixture at unitarity.

As mentioned above,
anti-de Sitter space-time does not obey this so-called Lifshitz
scaling but instead  a relativistic scaling with $z=1$.
Therefore, the usual AdS/CFT correspondence has to be modified to
be useful for this purpose.  One possibility is to geometrically
realize the full Schr\"odinger symmetry of the quantum critical
point as in \cite{Son:2008ye,Balasubramanian:2008dm} which, however, 
has been shown in \cite{thermoSchro1,thermoSchro2} to lead to different thermodynamical properties than fermions at unitarity. 
Another way to study the duality with an arbitrary dynamical exponent $z$
is to consider a simple generalization of the anti-de Sitter space-time that exhibits the appropriate dynamical scaling and is now known as the
Lifshitz space-time \cite{Kachru}. Here, particle-hole symmetry is still an exact property of the conformal field theory on the boundary, just like in the anti-de Sitter case. This is the background that we use in this work\footnote{These backgrounds also exhibit subtleties as they are not geodesically complete and give rise to diverging tidal forces at the origin of the spacetime, see \cite{Horowitz} for a very recent discussion. We will avoid these issues in this work by regulating the theory by turning on an infinitesimal black-hole horizon around the singularity.}.


\subsection{Single-particle correlators and ARPES sum-rules}

From the condensed-matter point of view we are  interested in
calculating the {\it single-particle} correlation function, or Green's
function, which for electronic systems is  directly
observable by angle-resolved photo-emission spectroscopy (ARPES),
or for ultracold atoms by radio-frequency (RF) spectroscopy. In particular, the 
{\it single-particle} Green's function $G({\vec k},\omega)$ differs from the retarded correlation function of a {\it composite operator} 
in that the former obeys a special sum-rule which is a direct consequence of the canonical (anti)commutation relations 
of the elementary fields. It reads
\begin{equation}\lab{sumint}
 \frac{1}{\pi} \int_{-\infty}^{+\infty} {\rm d}\omega\,  {\rm Im}\hspace{1pt} \bigl[G({\vec k},\omega)\bigr] = 1\ .
\end{equation}
A bulk fermion field $\Psi$ in the AdS/CFT correspondence would typically correspond to 
composite operators, schematically  of the form $\tr\le(\phi\cdots \phi \psi\phi\cdots\phi\ri)$ where $\phi$ and $\psi$ represent bosonic and fermionic elementary fields, respectively, in the adjoint representation of the dual gauge theory. These operators 
 satisfy modified sum-rules that differ from (\ref{sumint}), see \cite{Herzog} for a recent discussion. 
Intuitively it is clear that (\ref{sumint}) will not be obeyed because of the UV divergence that arises from placing more than one operator at the 
same point in space-time. 

To see precisely how the elementary sum-rule fails with the standard prescription of the AdS/CFT correspondence, it is important to realize that the Dirac equation in the $(d+1)$-dimensional bulk contains an undetermined mass parameter $M$ that is related to the scaling dimension of the dual fermionic operator. If we consider the case of $d$ even, then we can split any fermionic
operator into its chiral components ${\cal O}_\pm$. The scaling dimensions for these operators are then given by 
 \begin{equation}\label{scadim0}
\Delta_{\mp} = \frac{d+z-1}{2} \mp M \ .
\end{equation}
 From the AdS/CFT correspondence we 
obtain, as shown explicitly in this paper, that the Green's function of the operator ${\cal O}_-$
satisfies $G_{{\cal O}_-}({\vec 0},\omega) \propto \omega^{2M/z}$ at zero
temperature and at momentum ${\vec k} = {\vec 0}$. 
With this form of the Green's function, the only way to satisfy the requirement (\ref{sumint})  
in the usual AdS/CFT correspondence is to choose the mass of the bulk fermion dual 
to the operator $\cO_-$ as 
$M=-z/2$, in the appropriate units set by the radius of the Lifshitz background. Naively this would correspond to 
the trivial case of a free theory. 
However, a detailed analysis, 
which is somewhat more difficult in the special case
$M=-z/2$, shows that there are logarithmic corrections that yield a Green's function that obeys the exact Dyson equation\begin{equation}
G({\vec 0},\omega) = \frac{1}{\omega-\Sigma({\vec
0},\omega)}~,
\end{equation}
with a self-energy of the form $\Sigma(\vec{0},\omega)\sim \omega \log\omega$.  
This situation indeed satisfies the zeroth frequency sum rule, but instead it spoils the causality of
the theory that is signaled by a violation of the Kramers-Kronig
relation. Put differently, the logarithmic corrections to the
self-energy lead to the appearance of a (Landau) pole in the upper
half of the complex frequency plane, which is physically
unacceptable. 

It therefore appears that the standard prescription of
the AdS/CFT correspondence does not allow for the calculation of
the single-particle Green's function that is of crucial importance
in condensed-matter physics as well as 
 for strongly correlated systems. As mentioned above, this is of course expected, as the standard AdS/CFT correspondence calculates Green's function of 
 composite operators, not of the single-particle excitations that satisfy canonical (anti-)commutation relations\footnote{As an example of such an elementary field, one can consider the ${\cal N}=4$ super Yang-Mills theory with gauge group $U(N)$ and think of the trace of one of the four canonical Dirac fermions in the theory. In the $SU(N)$ theory this would be trivial, whereas it yields an elementary gauge-invariant fermionic operator in the $U(N)$ theory.}. 
  In view of this situation, the main
aim of this paper is to point out a modified prescription that does
allow for the extraction of a single-particle Green's function
that obeys both the zeroth frequency sum rule and the
Kramers-Kronig relation. We will find that this prescription can in fact be applied
when $-z/2<M < z/2$.

The key ingredient of our construction
is to introduce a UV cut-off surface close to the boundary, on which the dynamics of the elementary fermion interacting with the CFT
is defined. This introduces a UV scale such that, at zero temperature and at zero spatial momenta, 
there is still a dimensionful scale at the critical point that allows for non-trivial single-particle Green's functions. The idea is to apply the {\em alternative quantization} where one integrates over the boundary value of the bulk field, on the UV cut-off surface in the presence of a kinetic term for this bulk value, instead of fixing it to be a stationary source. This procedure computes the two-point function of the dynamical source, which we propose to correspond to the elementary fermion in the corresponding field theory.   

As we will show, the new prescription leads to the following retarded single-particle Green's function for a fermion coupled to the CFT at vanishing spatial momenta, temperature and chemical potential
\begin{equation}\lab{Giste}
G(\omega)=\frac{1}{\omega + {\tilde g}\, G_{{\cal O}_-}(\omega)}\ .
\end{equation}
Here, ${\tilde g}$ is a dimensionful coupling constant related to the position of the UV-brane. We will show that with this prescription both ARPES sum rules and Kramers-Kronig relations are satisfied.

The aforementioned restriction $-z/2<M < z/2$ can easily be understood from the field theory point of view. In view of (\ref{scadim0}) we observe that $M=-z/2$ corresponds to the unitarity bound for the operator $\cO_-$ of dimension $\Delta_+$. Below this value, the expression (\ref{Giste}) has singularities in the upper half plane, thus both the Kramers-Kronig relations and the sum-rule fails. On the other hand, recalling that  $G_{{\cal O}_-}(\omega)\propto \omega^{2M/z}$, one finds that, when $M>z/2$ the interaction term becomes irrelevant in the IR, but instead it modifies the UV. Therefore the sum-rule hence the unitarity fails (although one can satisfy Kramers-Kronig relations with an appropriate sign of $g$ in this case). This restricts the range of the operator coupled to the elementary field to lie between  $-z/2<M < z/2$. We find the corresponding restrictions on the gravitational side. As usual, the lower bound $M>-z/2$ corresponds to the requirement of finite energy for the bulk fluctuation corresponding to the operator $\cO$. The upper bound $M<z/2$  on the other hand, corresponds to requiring that the bulk field is normalizable near the boundary. If $M>z/2$, one cannot apply the aforementioned {\em alternative quantization}, instead one has to use the regular quantization, which would yield the inverse of (\ref{Giste}).    

The introduction of extra elementary degrees of freedom in this manner is not  novel: Contino and Pomarol \cite{Pomarol} considered the possibility of having a kinetic term for a fermion on a cut-off surface, hence a {\em dynamical source} field,  and have analyzed its consequences for the dual gauge theory in the context  of beyond-the-standard-model physics. However, our prescription in section \ref{prescription} differs from \cite{Pomarol} in that, although we stay within the unitarity bound, we are able to generate a Green's function of the form (\ref{Giste}) by considering the Green's function of the dynamical source instead of the operator $\cO_-$ that couples to it.  Perhaps the most recent and relevant work that has considered introduction of elementary fermions in the context of AdS/CFT is \cite{FP}, building upon the ideas in \cite{HTPS}. In this ``semi-holographic" method, an additional elementary fermion field is introduced and coupled to an operator $\cO_-$ in the CFT. The Green's function of the elementary field is obtained by field-theory methods, in terms of the correlator of $\cO_-$, which results in an expression similar to (\ref{Giste}). In a sense, the method that we advocate in this paper can be viewed as a {\em derivation of the semi-holographic method}. We propose a derivation of the expression in \eqref{Giste} by holographic methods, with no ad hoc introduction of an extra degree of freedom. In this paper, we focus here on the more universal sector with zero chemical potential and leave the generalization to nonzero chemical potential to future work.

The remainder of this work is organized as follows. In section 2 we 
study a Lifshitz background in the presence of a neutral black-brane
and derive the corresponding Dirac equation. Furthermore, we present our prescription to compute the Green's function for elementary fermions, for all values of the critical exponent $z$.
In section 3, we present and analyze the results of the above-mentioned prescription. Note that since we are considering a neutral black-brane, we describe the  normal, but quantum critical, state of the
system at zero chemical potential where particle-hole symmetry
is exact. 
Finally, we end in section 4  with some
conclusions and an outlook for future work.


\section{Generalities}


\subsection{Lifshitz backgrounds}\label{sect:prelim}

We start by considering  the action for a gravity system coupled to a gauge field with field strength $F_{\mu\nu}$ and a scalar field $\phi$ in $d+1$ space-time dimensions,
\eq{
  \label{action_01}
  S_0 = \frac{1}{16\pi G_{d+1}} \int {\rm d}^{d+1} x \:\sqrt{-g}\: \biggl[ \: R - \Lambda - \frac{1}{2} \:
  (\partial_{\mu} \phi )   (\partial^{\mu} \phi ) - \frac{1}{4} \: e^{\lambda \phi}  F_{\mu\nu} F^{\mu\nu}\:
  \biggr] \;.
}
The conventions for the space-time signature is $(-+\cdots +)$, and the sign of the cosmological constant $\Lambda$ is negative for negatively curved space-times. Moreover, we use units in which $\hbar=c=1$ and $G_{d+1}$ is the dimensionless 
Newton coupling constant. All length scales are measured in terms of the AdS or Lifshitz radius.

A black-brane solution that asymptotes to Lifshitz spacetime  is given by \cite{Taylor}
\eq{
  \label{metric_02}
  {\rm d}s^2 = \frac{{\rm d}r^2}{r^2\, V^2(r)} - V^2(r) \, r^{\,2z} {\rm d}t^2
  + r^{\,2} {\rm d} {\vec x}^{\:2} \;,\hspace{40pt}
  V^2(r) = 1 - \Big(\frac{r_h}{r}\Big)^{\,d+z-1} \;.
}
The radial coordinate $r$ runs from the boundary at $r=\infty$ to the
horizon of the black brane at $r=r_h$, and the temperature of the black brane is obtained by demanding the absence of a conical singularity at $r_h$ which leads to
\eq{
\lab{T}
T = \frac{d+z-1}{4\pi} \: (r_h)^{z} \;.
}
Note 
that the units of temperature are such that Boltzmann's constant  $k_B=1$ and 
that the metric \eqref{metric_02} enjoys the following Lifshitz isometry
\be\lab{scaling}
r\to \l r\;, \qquad \quad
t\to \l^{-z} t\;, \qquad \quad
x\to \l^{-1} x\;,  \qquad \quad
T \to \l^z T\;.
\ee
For the field strength $F_{\mu\nu}$ and the scalar field $\phi$, the solution to the equation of motion is given in terms of an arbitrary constant $f$
\eq{
  \label{sol_01}
  F_{rt} =\displaystyle f\, r^{\,z + d-2} \; , \hspace{40pt}
  e^{\lambda \phi} = \displaystyle 2(z-1)(z+d-1)\,\frac{r^{\,2(1-d)}}{f^2}\; .
}
Furthermore, the dilaton exponent $\lambda$ and cosmological constant $\Lambda$ are fixed to be
\eq{
  \lambda = -{\sqrt{ \frac{2(d-1)}{z-1}}} \; ,\hspace{40pt}
  \Lambda = -(z+d-1)(z+d-2)   \;.
}
Notice that in the limit $z\rightarrow 1$,  the scalar field decouples. The theory then reduces to the neutral black-brane solution in anti-de Sitter space-time with isotropic scaling properties \cite{Taylor,Tarrio:2011de}. Hence our results also allow for the case $z=1$. For $z>1$, note that  the dilaton diverges at the boundary, and also the field strength diverges in such a way that a charge and chemical potential cannot be defined. To introduce a chemical potential, another gauge field needs to be considered whose solution has proper boundary conditions \cite{Tarrio:2011de}. For the purpose of the present paper, namely a study of spectral-density functions at zero chemical potential, this is however not needed.

The diverging asymptotic behavior is thought to be a problem for holography and a proper treatment of the holographic renormalization procedure is therefore needed. This has been initiated in \cite{arXiv:1107.4451,arXiv:1107.5562,arXiv:1107.5792}, though not for  Lifshitz space-times supported by a scalar field like in the present paper. 
However, this problem is not severe for the applications we have in mind for the following reason. The fermions that we introduce below do not couple to the dilaton and are not charged under the gauge field. They only couple to the metric, which has a good asymptotic behavior. Therefore, the fermionic boundary Green's functions we obtain in this paper are well defined, as we  show explicitly.


\subsection{Prescription to calculate Green's functions for elementary fermions}
\lab{prescription}

In order to study fermions in the dual boundary theory, we add  the following action to the gravity system given by \eqref{action_01},
\eq{
  \label{action_05}
  S_f[\Psi] = i\hspace{1pt}g_f \int {\rm d}^{d+1}x \sqrt{-g} \: \left(
  \frac{1}{2}\, \ov \Psi \overrightarrow{ \slashed{\mathcal{D}}} \Psi
  -  \frac{1}{2}\, \ov \Psi \overleftarrow{ \slashed{\mathcal{D}}} \Psi
  -
  M \,\ov
  \Psi \Psi \right)  + S_{\6}[\Psi] \;.
}
Our notation here is as usual: $\ov \Psi = \Psi^{\dagger} \Gamma^{\ul 0}$, $\slashed{\mathcal D} = \Gamma^{\ul a} e_{\ul a}{}^{\mu} \mathcal D_{\mu}$ with $\mathcal D_{\mu} = \partial_{\mu} + \frac{1}{4} \omega_{\mu \ul a \ul b} \Gamma^{\ul a \ul b}$, and $\Gamma^{\ul a}$ are the $(d+1)$-dimensional gamma matrices, where $\Gamma^{\ul 0}$ is antihermitean in our conventions.
Furthermore, $e_{\ul a}{}^{\mu}$ are the vielbeins and $\omega_{\mu}{}_{\ul a \ul b}$ is the spin connection, for which indices are raised and lowered with the flat metric $\eta_{\ul a\ul b}={\rm diag}\,(-1,+1,\ldots,+1)$. Finally, we defined $\Gamma^{\ul a \ul b} = \half[\Gamma^{\ul a},\Gamma^{\ul b}]$. Next, since we are interested in a space-time with boundaries, we introduce an action $S_{\6}[\Psi]$ to make the variational problem well-posed \cite{FerCor, Pomarol}.
More concretely, we  consider boundaries  at the horizon $r=r_h$ as well as at a cut-off surface located at $r=r_0$. But, as we will see below, due to the fact that $\sqrt{g^{rr}}$ vanishes at the horizon we only need to add a boundary term to the action for the cut-off surface at $r=r_0$.

Next, let  us define  components $\Psi_{\pm}$ of the Dirac fermion $\Psi$ using the gamma matrix $\Gamma^{\ul r}$ as follows
\eq{\lab{chiral}
  \Psi_{\pm} \equiv \half\,\bigl(1\pm \Gamma^{\ul r}\bigr)\,\Psi\;,
  \hspace{60pt}
  \Gamma^{\ul r}\,\Psi_\pm=\pm \,\Psi_\pm \;,
}
where  for  $d$ being odd $\Gamma^{\ul r}$ is the radial gamma matrix, while for $d$ being even it is  the boundary chirality operator. In both cases, we choose it to be hermitian.
Note  that  a Dirichlet boundary condition for either $\Psi_+$ or $\Psi_-$  can be imposed consistently, however, one cannot impose both  $\delta\Psi_+=0$ and $\delta\Psi_-=0$ at the same time because the Dirac equation is of first order in derivatives. Mixed boundary conditions are also possible, and for the case when Lorentz symmetry  is broken on the boundary see \cite{Laia:2011wf}. Here, we focus on pure Dirichlet conditions for which the appropriate boundary action  reads
\eq{\lab{bndry}
  S_\6[\Psi] =
  \pm i\: \frac{g_f}{2}  \int_{r=r_0} {\rm d}^{d}x \sqrt{-h} \sqrt{g^{rr}} \le(\ov{\Psi}_-\Psi_+ +\ov{\Psi}_+
  \Psi_-\ri)\;,
}
where $h$ is the determinant of the induced metric on the boundary, and where the upper signs corresponds to imposing $\delta \Psi_+=0$ at $r=r_0$, while the lower sign corresponds to requiring $\delta \Psi_-=0$.
Which of the two Dirichlet conditions is imposed is a matter of convention because they can be related by inverting the mass  $M\to -M$ \cite{Liu2}. For definiteness, here we  make the choice
\eq{\lab{choice}
  \delta \Psi_+ = 0 \hspace{70pt} \mbox{at} \hspace{30pt}
  r=r_0\;.
}

Now, note that since we impose \eqref{choice} it is possible to add a ``UV action''  $S_{UV}[\Psi_+]$ on the cut-off boundary which does not obstruct the variational problem. If $d$ is even, this boundary theory describes a chiral spinor, whereas for $d$ being odd $\Psi_+$ is a Dirac spinor. In particular, as in \cite{Pomarol}  we shall introduce a kinetic term for the field $\Psi_+$ and hence make it dynamical 
\be
\lab{SUV}
S_{UV}[\Psi_+] = Z\int_{r=r_0} {\rm d}^dx\:\sqrt{-h}\: {\ol \Psi}_+ \slashed{D}_z(r,x) \Psi_+ \;,
\ee
where $Z$ is an arbitrary constant and $\slashed D_z(r,x)$ is the kinetic operator for a theory with dynamical exponent $z$.  
With $\Gamma^{\ul a}$ the appropriate Dirac matrices in $d$ dimensions, that is $\ul a = \{\ul t, \ul x_1, \ldots, \ul x_{d-1}\}$, this operator reads\footnote{The sums over the indices here are, a priori, over $d+1$ dimensions, including the radial direction. However the latter gives vanishing contribution as ${\ol \Psi}_+ \Gamma^{\ul r} \Psi_+ =0$ by bulk chirality.}  
\be\lab{Delta}
\slashed D_z(r,x) = i \hspace{1pt}\slashed{\6} = i\hspace{1pt} \Gamma^{\ul a}e_{\ul a}{}^\m  \6_\m\ ,
\ee
which becomes the usual kinetic term for a chiral fermion in the relativistic case $z=1$.
The addition of a spin-connection $\omega_{\mu}{}_{\ul a\ul b}$ on the brane is left out. In general, this would lead to mass terms, but for even $d$ the spinors are chiral and the term with the spin-connection vanishes. For odd dimensions, this can lead to Dirac mass terms, but we will not consider this possibility here.

This is how we introduce an elementary fermion in the theory. Its free dynamics is described by the boundary action (\ref{SUV}) and the contribution from coupling this excitation to the strongly coupled CFT is described by the general relativity (GR) action (\ref{action_05}) and (\ref{bndry}), evaluated on-shell. The total action 
\be\lab{totac}
S_{full}[\Psi_+]  = S_f[\Psi] + S_{UV}[\Psi_+] \;,
\ee
evaluated on shell then determines the two-point function of the elementary fer\-mi\-on field $\Psi_+$, where we eliminated $\Psi_-$ in favor of $\Psi_+$ via the Dirac equation.
Note that we cannot add a tree level mass term on the UV brane because this would also require $\delta \Psi_-=0$ for consistency of the variational principle. However,  a boundary mass for the chiral fermions can be introduced through a Higgs mechanism. Finally, we observe that a dilaton coupling in the UV action can be included, especially in view of the fact that the dilaton indeed couples to the brane fields with a factor $\exp(-\phi)$. But, the value of the dilaton at a fixed radial position $r=r_0$ can be absorbed into the constant $Z$ in (\ref{SUV}).

We also have to point out an important subtlety which has a well-known analog in the case of bosonic real-time correlation functions \cite{Son1}. The setting above would correctly give rise to an effective action for $\Psi_+$ {\em only for Euclidean signature}. To see this  subtlety, let us ignore the UV action (\ref{SUV}) for the moment and set $Z=0$. Then the prescription where we keep both terms in (\ref{bndry}) would produce a Green's function for the operator that couples to the source $\Psi_+$ which is not of the correct form in general. In particular, it would yield a {\em real} expression that is not generally true for real-time correlators. The way to circumvent this problem is well-known: to compute the Green's function we should ignore the first term $\ov{\Psi}_-\Psi_+$
in (\ref{bndry}). Namely we should replace (\ref{bndry}) by
\eq{
\lab{bndry1} 
S_\6 =  i \, g_f
\int_{r=r_0} {\rm d}^{d}x \sqrt{-h} \sqrt{g^{rr}}\: \ov{\Psi}_+\Psi_-\;,
}
where we included an additional factor of two to be consistent with the usual conventions.
Equivalently, we can also use the prescription provided in \cite{Liu1,Liu2}.

\bigskip
Let us now become more concrete and specify the boundary space-time dimensions to be $d=4$. Also, with slight abuse of notation, we write the four-component  Dirac spinor $\Psi$ in terms of two-component spinors $\Psi_+$ and $\Psi_-$ as
\begin{equation}
\Psi=\begin{pmatrix}\Psi_+ \\ \Psi_-\end{pmatrix} \;.
\end{equation}
The four-dimensional gamma matrices can be expressed in terms of $\sigma^{\ul a} = (1, \vec\sigma)$ and $\ov\sigma^{\ul a} = (-1, \vec\sigma)$ with $\sigma^{\ul i}$ being the Pauli matrices in the following way
\eq{
  \Gamma^{\ul a} = \left( \begin{array}{cc} 0 & \ov\sigma^{\ul a} \\ \sigma^{\ul a} & 0 \end{array} \right)
  \;.
}
The actions \eqref{SUV} and \eqref{bndry1} then become
\begin{align}
\lab{SUV_spec}
& S_{UV}[\Psi_+] = - Z \int_{r=r_0} {\rm d}^dx\:\sqrt{-h}\:  \Psi^{\dagger}_+ \slashed D_z(r,x) \Psi_+ \;, 
\hspace{25pt} \slashed D_z(r,x) =  i \sigma^{\ul a} e_{\ul a}{}^\m  \6_\m  \;, \\
\lab{bndry1_spec}
&S_\6 =  - i \, g_f
\int_{r=r_0} {\rm d}^{d}x \sqrt{-h} \sqrt{g^{rr}}\:  \Psi^{\dagger}_-\Psi_+\;,
\end{align}
where again $\ul a = \{\ul t, \ul x_1, \ldots, \ul x_{d-1}\}$.
The two-component spinors $\Psi_-$ and $\Psi_+$ are not independent but are related through the boundary condition at the horizon of the black brane, which should be chosen as the {\em in-falling} one for the {\em retarded} Green's function.
Following \cite{Liu1}, we define Fourier-transformed spinors on each constant $r$-slice as
\begin{equation}
\Psi_{\pm}(r,x)=\int \frac{{\rm d}^dp}{(2\pi)^d}\,\psi_{\pm}(r,p)\,e^{ip_\mu x^\m}\; ,
\hspace{40pt} p_\mu=(-\omega,\vec k)\ .
\end{equation}
The Dirac equation then imposes a relation between the chiral components $\psi_+$ and $\psi_-$ of the form
\eq{
\lab{xi1} 
\psi_-(r,p) = -i \hspace{1pt}\xi(r,p)\, \psi_+(r,p)\ ,
}
where $\xi$ is a two-by-two matrix. This matrix yields the Green's function of the operator $\cO_-$ in the CFT that couples to the source $\psi_+(p)$ when evaluated on the boundary \cite{Liu1} 
\eq{
\lab{GO}
G_{\cO_-}(p) = - \lim_{r\to r_0} r^{2M} \xi(r,p)\;,
}
where $G_{\cO_-}$ denotes the Fourier transform of the Green's function which takes the form $i \la\{ \cO_-(x), \cO_-^{*}(x')\}\ra \Theta(t-t')$.

Next, substituting  \eqref{xi1} into the Fourier transforms of \eqref{SUV_spec} and \eqref{bndry1_spec}, we find for the on-shell action \eqref{totac} that 
\be
\lab{totac1}
S_{full}[\Psi_+]  = - \int_{r=r_0}\frac{{\rm d}^dp}{(2\pi)^d}\:\sqrt{-h}\:  \psi_+^\dagger \le[
  g_f \sqrt{g^{rr}} \xi (r,p) +  Z \slashed D_z(p) \ri] \psi_+\;,
\ee
where $\slashed D_z(p)= - \sigma^{\ul a} e_{\ul a}{}^{\mu} p_{\mu}$.
We regard this expression as the {\em effective action of the elementary field $\Psi_+$} that is coupled to an operator $\cO_-$ in the CFT with scaling dimension
\be\lab{scadim}
\Delta_+ = \frac{d+z-1}{2} + M\ ,
\ee
consistent with \eqref{scadim0}. The convention here is to write $\Delta_+$ for an operator ${\cal O}_-$ that couples to $\Psi_+$.
The effect of this coupling is described  by the first term in (\ref{totac1}). Next, to obtain the Green's function of a canonically normalized  field, we transform $\Psi_+\to \Psi_+ Z^{-1/2} r_0^{(1-d)/2}$ and find the Green's function of the elementary field as
\be\lab{GR1}
G_R(r_0,p) = - \le(r_0^z V(r_0) \slashed D_z(p) + \frac{g_f}{Z} r_0^{1+z}V^2(r_0)\: \xi(r_0,p) \ri)^{-1}\ ,
\ee
where we remind that the blackness function $V(r)$ appears in the metric given by (\ref{metric_02}). In the limit of zero temperature, or for temperatures small compared to the UV cut-off,
we have $r_0/r_h\gg 1$ so that we can approximate $V(r_0)\simeq 1$. Once (\ref{GR1}) is obtained in this way, we take a {\em double scaling limit} 
\be\lab{dslim} 
r_0\to\infty, \qquad g_f\to 0, \qquad g_f r_0^{1+z-2M} = {\rm const.}
\ee
Furthermore, we note the following consistency checks: 
\begin{itemize}

\item[(i)] In the case of $g_f=0$ and finite $r_0$, hence the dynamical source field is decoupled from the CFT, we find the correct propagator for the free field $\Psi_+$.

\item[(ii)]  In the case of non-vanishing $Z$ we calculate the correlation function by taking functional derivatives of the generating function $\mathcal Z[J_-]$ where $J_-$ is the {\em Legendre transform} of $\Psi_+$. Schematically, this reads   
\eq{
\lab{altquant} 
G_R \propto \frac{\delta^2 \mathcal Z[J_-]}{\delta J_-^\dagger  \delta J_-}
\hspace{11pt}   {\rm with} \hspace{11pt}
\mathcal Z[J_-] = \int {\cal D} \Psi_+ e^{ i S_{full}[\Psi_+] + i \int {\rm d}^dx( {\Psi}_+^\dagger J_- + {J}_-^\dagger \Psi_+ ) },
}
where $S_{full}$ is given by (\ref{totac1}). This provides an alternative way to arrive at the result (\ref{GR1}). At the end of this calculation one takes the limit (\ref{dslim}) as explained above.

\item[(iii)] In the case of $Z=0$, that is the UV dynamics for the source $\Psi_+$ is turned off, and
after an appropriately rescaling $\Psi_+\to Z^{\half} \Psi_+$  we obtain the correct result for the CFT operator $\cO_-$ {\em in the alternative quantization},\footnote{We thank Hong Liu for a discussion on this issue.} that is the inverse of the result in the standard quantization. This is expected because in the {\em alternative quantization} above, we integrate over the source, see \cite{Klebanov:1999tb}, rather than fixing it.
\end{itemize}

In passing, let us comment on the range of the parameter $M$ for general $z$.  In analogy with the AdS case, $z=1$, we find that, among the possible  boundary asymptotics of $\Psi$, $\Psi_+$ is non-normalizable near the boundary for $M>z/2$. Therefore it should correspond to the source term in this range. On the other hand, in the range $0\leq M < z/2$ both $\Psi_+$ and $\Psi_-$ are normalizable, therefore one has the possibility of choosing either of them as the source. This corresponds to two different quantizations \cite{Klebanov:1999tb}. Using the fact that the system is invariant under the exchange $\Psi_\pm \to \Psi_\mp$ and $M\to -M$, one instead extends the range of $M$ to $M>-z/2$ and always fixes $\Psi_+$ to be the source. In this way, one covers all  possible quantizations. In particular, the range $-z/2<M< 0$ now corresponds to the alternative quantization where $\Psi_-$ is chosen as the source. This is what we shall do in the following, i.e., we will always choose $\Psi_+$ as the source but consider the wider range  $M>-z/2$.

As in the AdS case \cite{Liu2}, we can easily derive a first-order differential equation that determines the matrix $\xi$. For this purpose, let us specify to the case of $d=4$ for which $\Psi_+$ describes a two-component spinor.  A similar analysis can be done for $d=3$, as long as $z<d-1$.
The up and down components  $u_{\pm}$ and $d_{\pm}$ of $\Psi_{\pm}$ are defined as follows
\eq{\lab{ud}
\psi_\pm (r,p)= V(r)^{-\half} r^{-\half(d+z-1)}\binom{u_\pm (r,p)}{d_\pm (r,p)}\ ,}
where the overall $r$-dependent factor is introduced to simplify the fluctuation equations.
Then, the eigenvalues of the matrix $\xi$ above are given by the ratios
\eq{\lab{defxi} \xi_+ = i\frac{u_-}{u_+}\ , \hspace{60pt} \xi_-
= i\frac{d_-}{d_+}\ .
}
Employing the rotational symmetry of the problem, we can choose a basis where the particle moves along the $z$-axis, i.e. $\vec{k} = (0,0,k)$.
Then we first derive the following first-order decoupled equations, that relate the plus and minus
components in \eqref{ud} as
\eq{
  \lab{des}
  \arraycolsep2pt
  \begin{array}{lcllcl}
  i(\tilde{\omega} +k) u_+ &=& \mathcal A(-M)u_-\; ,\hspace{60pt} &
  i(\tilde{\omega} -k) u_-& = & \mathcal A(M)u_+\; ,  \\[1.7mm]
  i(\tilde{\omega} -k) d_+ &=& \mathcal A(-M)d_-\;,\hspace{60pt} &
  i(\tilde{\omega} +k) d_- &=& \mathcal A(M)d_+\; ,
  \end{array}
}
where we introduced the shorthand notations
\be \lab{shorthand}
\tilde{\omega} =  -\frac{\omega}{r^{z-1}V}, \hspace{40pt}
\mathcal A(M)\equiv r\bigl(r V \6_r - M \bigr)\; .
\ee
Using the first-order equations above we can now
derive a {\em first-order equation} for  $\xi_+$ and
$\xi_-$ defined in (\ref{defxi}) as 
\eq{\lab{xieq} 
r^2V \partial_r \xi_\pm + 2Mr \xi_\pm = -\tilde{\omega} \mp k + (-\tilde{\omega} \pm k)
\xi_\pm^2 \;.
}
The in-falling boundary conditions at the horizon correspond to
\eq{
\lab{bc} 
\xi_\pm(r_h,p) = i\;.
}
Once the function $\xi_\pm$ is obtained, the Green's function for the elementary field coupled to the CFT is given by the equation (\ref{GR1}) above.
In passing, let us  also observe that by a  change of variables we find that  $\xi_{\pm}(r,\omega, k)$ can be expressed in terms of the following ratios
\eq{
   \xi_{\pm}(r, \omega, k ) 
   = \xi_{\pm} \left( \frac{r}{k}, \frac{\omega}{k^z} \right)
   =\xi_{\pm} \left( \frac{r}{\omega^{1/z}}, \frac{k}{\omega^{1/z}} \right) \;.
}


\section{Results}


\subsection{The relativistic CFT with $z=1$}

In the case that the elementary fermion $\Psi_+$ is coupled to an operator $\cO_-$ in a {\em relativistic} CFT, the background is given by an AdS black brane, the case $z=1$ in (\ref{metric_02}). For simplicity of the notation, we choose again the two-component spinors introduced in \eqref{ud}. The kinetic term (\ref{Delta}) in the zero-temperature limit\footnote{As noted in the introduction, we obtain the zero $T$ result, by first performing the calculation at infinitesimally small horizon and then take the limit $T\to0$.} and for $z=1$, then takes the form
\begin{equation}
 r_0 \slashed D_1(r_0,\omega,{\vec k}) =
  \omega -\vec \sigma\cdot \vec{k} \ .
  \end{equation}
Let us now focus on the case $\vec k=0$. The solution to the differential equation \eqref{xieq} with infalling boundary conditions $\xi_{\pm}(r=0)=+i$ reads
\eq{
  \xi(r,\omega) = \frac{J_{M-\frac12} \left( \frac{\omega}{r} \right) + e^{i\pi (M+\frac12)} 
  J_{-M+\frac12} \left( \frac{\omega}{r} \right)}{
  J_{M+\frac12} \left( \frac{\omega}{r} \right) - e^{i\pi (M+\frac12)} 
  J_{-M-\frac12} \left( \frac{\omega}{r} \right)
  } \;,
}
where $J_{\alpha}(x)$ are Bessel functions of the first kind. 
Focussing on the case $-1/2<M<+1/2$, performing  an expansion for large $r$ and keeping only the leading order term, we find
 the following expression \cite{Liu2} 
\eq{\lab{AdSKT}
\xi(r_0,\omega) \simeq 
- \le(2r_0\ri)^{-2M} \frac{\Gamma\le(\half-M\ri)}{\Gamma\le(\half+M\ri)}e^{-i\pi (M+\half)} \omega^{2M}
\;.
}
The bulk mass $M$ and the scale dimension of the operator $\cO_-$ is given by $\Delta_+ = d/2 + M$. From (\ref{GR1}) we then find the Green's function for the up (+) and down (-) components of the spinor $\Psi_+$,
which reads 
\be\lab{GRAdS1}
G_{R}(\vec 0, \omega) = - \frac{1}{\omega -  \tilde{g}_M\:  \omega^{2M} e^{-i \pi (M+\half)}} \ ,
\ee
where we remind the reader that in \eqref{GRAdS1} a two-by-two identity matrix is understood, and
where we defined the constant
\be\lab{gtilde}
\tilde{g}_M =  \frac{g_f}{Z}\, 2^{-2M} \frac{\Gamma\le(\half-M\ri)}{\Gamma\le(\half+M\ri)} r_0^{-2M+2}.
\ee
Note that in case $\tilde{g}_M$ vanishes, hence $\Psi_+$ is decoupled from the CFT, one should add a factor of $i \eps$ with $\eps>0$ in the denominator of \eqref{GRAdS1}, as usual for the retarded  Green's function. Furthermore, $\tilde{g}_M$ is positive definite in the range $-1/2 < M <+1/2$ provided $g_f/Z$ is positive definite.

As can be seen from \eqref{GRAdS1}, the self-energy is proportional to $\omega^{2M}$ which for $M<0$  diverges in the infrared. This is an interesting situation with possible applications in condensed-matter physics. 
Note also that the retarded Green's function satisfies
\begin{equation}\lab{relimp}
G_R^\dagger (\vec 0, \omega)=-G_R(\vec 0 , -\omega)\;,
\end{equation}
which, using \eqref{xieq}, can be  generalizes to non-vanishing spatial momentum as
\begin{equation}\lab{parole}
\tr\, G_R^\dagger (\vec{k},\omega)=-\tr\, G_R(\vec{k},-\omega)\;.
\end{equation}

Now, from the expression (\ref{GRAdS1}) we find a pole at $\omega=0$, and a branch-cut that we can place on the negative imaginary axis.\footnote{There is a physical reason for this. As one turns on temperature the branch-cut disintegrates into an infinite series of poles which should appear on the negative imaginary axis \cite{MIT2}. We thank David Vegh for a discussion on this point.} 
The location of the only other possible pole in this expression 
depends on the sign of $\tilde g$. More concretely, 
in the range $-1/2<M<1/2$ there is no pole in the upper half-plane of the principal sheet provided that
\be
{\tilde g}_M >0 \ .
\ee
Thus, the expression (\ref{GRAdS1})
satisfies the Kramers-Kronig relations. Moreover, using Cauchy's theorem for a contour in the upper half plane with a semi circle closing at infinity, we find
\begin{equation}\label{contour-G}
\int_{-\infty}^\infty {\rm d}\omega \, \tr\, G_R(\vec{k},\omega)=2\pi i \ .
\end{equation}
For this identity, it is essential that $M<1/2$ such that at infinity the fall-off of the Green's function goes like $\omega^{-1}$, and such that the pole at $\omega=0$ does not contribute.
A sum rule now follows straightforwardly. Introducing the spectral density function\footnote{Note the relative minus sign in this definition with respect to the more conventional definition,
which we denoted by $G$ in the introduction. This is because we define our 
correlators with an overall minus sign, following the convention of \cite{Liu2}.}
\begin{equation}
\label{spec_dens}
\rho(\vec k,\omega) \equiv  \frac{1}{2\pi}\, {\rm Im}\,\, \tr\, [ G_R (\vec k,\omega)] \;,
\end{equation}
and employing \eqref{relimp} as well as \eqref{contour-G}, it then follows that\footnote{Employing \eqref{GRAdS1} for the Green's function in \eqref{spec_dens}, we also checked analytically that the sum rule is obeyed.}
\begin{equation}
\int_{-\infty}^\infty {\rm d}\omega \,\rho(\vec k,\omega)=1\;.
\end{equation} 
This is the ARPES sum rule required for one-particle fermionic states. It holds for any value $-1/2<M<1/2$.
The fact that the sum-rule is violated outside this range is in accord with the observation that $M=-1/2$ corresponds to the {\em unitarity  bound} for fermions in a relativistic CFT.

\bigskip
For non-vanishing momenta, one can easily redo the calculation. The non-trivial part of (\ref{GR1}) is the self-energy that is given in terms of $\xi(r,p)$. 
This is of course equivalent to knowing the two-point function of the operator $\cO_-$ in the CFT, which is a known result in the literature 
also for non-zero spatial momentum \cite{Sfetsos,Muck,hep-th/9902137,Liu2}. In the two-component notation, the result can be expressed as
\eq{
  \label{function_G}
  G_{\cO_-}(p)= T(p) \, \sigma^{\ul a} \,\delta_{\ul a}{}^{\mu} p_{\mu} \;,
}
where the function $T(p)$ for large values of $r$ is given by
\eq{
\label{function_T}
&T(p) \simeq (2r)^{-2M}\frac{\Gamma(\half-M)}{\Gamma(\half+M)} 
\times \\ 
&\hspace{30pt}\times\left\{\begin{array}{l@{\hspace{20pt}}l@{\hspace{20pt}}r@{}} 
+p^{2M-1}e^{-i\pi(M+\half)} \;, & p\equiv\sqrt{\omega^2-|\vec{k}|^2}\;, & \omega> +|\vec{k}| \;, \\
+p^{2M-1}e^{+i\pi(M+\half)}\;,  & p\equiv\sqrt{\omega^2-|\vec{k}|^2}\;, & \omega< -|\vec{k}|\;, \\
-p^{2M-1} \;, & p\equiv\sqrt{|\vec{k}|^2 - \omega^2}\;, & - |\vec{k}|<\omega< +|\vec{k}| \;.
\end{array} \right.
}
In order to investigate the analytic structure, we have to find the analytic continuation of the function $T(p)$ in the entire complex $\omega$ plane. We note that 
to obtain the second line in \eqref{function_T} with $\omega<-|\vec{k}|$ from the first with $\omega>+|\vec k|$, we take $p\to e^{i\pi} p$. Similarly, to obtain the last line we take $p\to e^{i\pi/2} p$.  Therefore, the single expression 
\be\lab{Tgen}
T(p) = (2r)^{-2M}\frac{\Gamma(\half-M)}{\Gamma(\half+M)} e^{-i\pi(M+\half)}p^{2M-1}\;,
\hspace{40pt} p\equiv\sqrt{\omega^2-|\vec{k}|^2} \;,
\ee
covers the entire complex $\omega$ plane provided we include a branch-cut that runs from $\omega=0$ to $\omega= -i \infty$ and take the first sheet to make it single-valued. So in particular, this means that $-1$ will be represented by $e^{+i\pi}$.  Now, the full Green's function (\ref{GR1}) is given by 
\begin{equation}\label{GRAdS2}
G_R(\vec k,\omega)=-\frac{1}{p^2\le(1-{\tilde g}_Me^{-i\pi(M+\half)}p^{2M-1}\ri)}
\Big(\omega+\vec \sigma \cdot \vec k\Big) \ ,
 \end{equation}
 where the constant ${\tilde g}_M$ is defined in (\ref{gtilde}). 
The poles of the retarded Green's function coming from the first term in the denominator of \eqref{GRAdS2} are at
\begin{equation}
 \omega=\pm k\; ,
\end{equation}
and there are no poles in the upper half-plane of the principal sheet if
\eq{
  {\tilde g}_M >0 \;.
}  
Thus, the Kramers-Kronig relations are obeyed. 
Also, 
using the symmetry property (\ref{parole}) and the fact that  the large frequency behavior is the same as for $k=0$, one finds again that \eqref{contour-G} is satisfied, and hence the sum rule is obeyed.\footnote{Using \eqref{function_G} and \eqref{function_T} in the expression for the full Green's function in \eqref{spec_dens}, we also checked analytically that the sum rule for non-vanishing $k$ is obeyed.}


\subsection{The Lifshitz case}

For  values of the dynamical exponent $z$ different from one, there are a number of important changes. We consider again the case of vanishing temperature, implying $V(r)=1$, and first assume both $k$ and $\omega$ to be non-zero. The kinetic term in \eqref{totac1}
then reads
\begin{equation}
 r_0^z \slashed D_z(\omega,{\vec k}) =
  \omega -r_0^{z-1}\vec \sigma\cdot \vec{k} \ .
\end{equation}
After rescaling $\Psi_+\to \Psi_+ Z^{-1/2} r_0^{(1-d)/2}$,  the action (\ref{totac1})  becomes 
\be\lab{totac2}
S_{full}[\Psi_+]  = - \int_{r=r_0} \frac{{\rm d}\omega\: {\rm d}^{d-1}k}{(2\pi)^d}\:  \psi_+^\dagger \le[
  \frac{g_f}{Z}\, r_0^{1+z} \xi (r,p) +  \omega - r_0^{z-1} \vec{\sigma}\cdot\vec{k} \ri] \psi_+\ .
\ee

Now, at the end of our calculation, we will remove the UV cut-off by taking the limit $r_0\to\infty$. Consequently, a possible divergence in the first term in (\ref{totac2}) 
should be absorbed in the redefinition of the parameter $g_f$. However, the divergence in the last term  should  be cancelled by a 
counter-term action, which is parallel to what happens in field theory. To explain that point, 
let us consider a single spin-component $\psi$ of a fermion in a Lifshitz invariant free theory.
The kinetic term reads $S_{kin} \propto \int   {\rm d}\omega\: {\rm d}^{d-1}k\: \psi^*(\omega + \eta k^z)\psi$.  The coupling $\eta$ is classically marginal, with the scaling $\omega \propto k^z$ and the classical scaling dimension for $\psi$ being $(d-1)/2$.  In presence of a UV cut-off in spatial momentum of order $\Lambda_k \propto r_0$, however, the Lifshitz scaling is broken and one has to consider the contribution of {\em relevant} terms. The kinetic action then becomes  $S_{kin} \propto \int   {\rm d}\omega\: {\rm d}^{d-1}k\: \psi^*(\omega + \eta k^z +\tilde{\eta} k)\psi$ where $\tilde{\eta}$ scales like $\tilde{\eta}\propto \Lambda_k^{z-1}$, hence the term is relevant in the IR limit. If we want to maintain Lifshitz scaling in the IR, we have to renormalize and remove $\tilde{\eta}k$, and we are  left only with a classically marginal spatial term $\eta k^z$. In the end of this procedure, from (\ref{totac2}) we obtain  the following Green's function 
\begin{equation}\label{GR2}
G_R(\vec k,\omega)= - \le(\omega + \eta\: \vec{\sigma}\cdot\vec{k}\: k^{z-1} +  \frac{g_f}{Z}\, r_0^{z+1}  \xi(r,p) \ri)^{-1} \;.
 \end{equation}

We also note that the matrix $\xi(r_0,p)$ in (\ref{GR2}) is not arbitrary but its form is determined by the scale dimension of the operator $\cO_-$ and the Lifshitz scaling. Choosing again a basis
where the particle moves along the $z$-axis, we can write $\vec{k} = (0,0,k)$ and obtain
 \be\lab{xibar}
 \lim_{r_0\to\infty} r_0^{2M} \xi(r_0,p)= k^{2M}\: \ov{\xi}_1\left(\frac{\omega}{k^z}\right) 
 = \omega^{\frac{2M}{z}} \:\ov{\xi}_2\left(\frac{\omega}{k^z}\right)\; .
\ee
Here we indicated explicitly that the matrices $\ov{\xi}_{1,2}$ are only a function of the ratio $\omega/k^z$. This is of course expected as it is essentially the Green's function of the operator $\cO_-$ in the Lifshitz CFT. We can see this in the holographic picture directly from equation \eqref{xieq}. First consider $T\to 0$  which means setting $V(r)=1$. Now we perform  a change of variables  $x=r/k$ under which \eqref{xieq} becomes
\eq{
\lab{xieqr} 
x^2 \frac{d}{dx} \xi_\pm(x) + 2M x  \xi_\pm(x) = \frac{\ov \omega}{x^{z-1}} \mp 1 + \left(\frac{\ov \omega}{x^{z-1}} 
\pm 1\right)\xi_\pm^2(x) \;,
}
where $\ov \omega=  \omega/k^z$. Note that $\omega$ and $k$ only appear in the combination $\ov \omega$, and the boundary condition (\ref{bc}) also respects this. Then, the scaling in \eqref{xibar} follows.

Unfortunately, for arbitrary $z$ we cannot solve (\ref{xieq}) analytically when both $\omega$ and $k$ are nonzero (an exception is the special case $M=0$ and $z=2$, see \cite{Korovin}). However, for the particular case $k=0$ and $\omega\ne 0$ the solution to (\ref{xieq}) for infalling boundary conditions \eqref{bc} reads
\eq{
  \xi(r,\omega) = \frac{
  J_{\frac Mz-\frac12} \left( \frac{\omega}{zr^z} \right) + e^{i\pi (\frac Mz+\frac12)} 
  J_{-\frac Mz+\frac12} \left( \frac{\omega}{zr^z} \right)}{
  J_{\frac Mz+\frac12} \left( \frac{\omega}{zr^z} \right) - e^{i\pi (\frac Mz+\frac12)} 
  J_{-\frac Mz-\frac12} \left( \frac{\omega}{zr^z} \right)
  } \;,
}
where $J_{\alpha}(x)$ are again Bessel functions of the first kind and where a two-by-two identity matrix is again understood.
Its  expansion for large values of $r$ can be obtained as
\eq{
\lab{anaw}  
  \xi(r,\omega) \simeq -r^{-2M} \frac{\Gamma(\half-\frac{M}{z})}{\Gamma(\half+\frac{M}{z})}
  e^{-i\pi \left( \frac12+\frac{M}{z}\right)}(2z)^{-\frac{2M}{z}} \omega^{\frac{2M}{z}}\;.
}
Note that these expressions hold for generic values of $M$, however for the special values  $M\in {\mathbb Z} + \half$  logarithmic terms appear. Furthermore, in the situation of $\omega=0$ and $k\neq0$ the solution to \eqref{xieq} reads as follows
\eq{
  &\xi_+(r,k) = \frac{
  I_{+M-\frac12} \left( \frac{k}{r} \right) +
  e^{2 \pi i (M+\frac12)} I_{-M+\frac12} \left( \frac{k}{r} \right)}{
  I_{-M-\frac12} \left( \frac{k}{r} \right) +
  e^{2 \pi i (M+\frac12)} I_{+M+\frac12} \left( \frac{k}{r} \right)
  } \;,  \hspace{50pt} k>0 \;,\\
  &\xi_+(r,k) = \frac{
  K_{+M-\frac12} \left( -\frac{k}{r} \right) }{
  K_{+M+\frac12} \left( -\frac{k}{r} \right) }  \;,
   \hspace{150pt} k<0 \;,
}
where $I_{\alpha}(x)$ and $K_{\alpha}(x)$ are the modified Bessel functions of the first and second kind, respectively.
Also, the infalling boundary conditions have to be modified to $\xi_{\pm}(0,k)=+1$, and the expansion for large values of $r$ in both cases reads
\eq{
\lab{anak}  
   \xi(r,k) \simeq - (2r)^{-2M} \frac{\Gamma(\half-M)}{\Gamma(\half+M)}~k^{2M-1}\; \vec k\cdot \vec \sigma
   \;.
}
Again, these results hold for generic $M$, but for $M\in z({\mathbb Z} +\half)$ logarithmic terms appear.

In order to investigate the analytic structure of the Green's function (\ref{GR2}), let  us consider the general case $\omega\ne 0\ne k$. Since we do not know the analytic expression for $\xi(r,p)$ in this case, we have to make the assumption that all possible non-analyticity of (\ref{GR2}) is in the lower half-plane, i.e., Kramers-Kronig relations are satisfied.\footnote{We study the analytic structure for this general case and at finite temperature {\em numerically} in \cite{upcoming}.} With this assumption, we can evaluate the integral 
\eq{
  \int_{-\infty}^{\infty} d\omega\, \rho(\vec k , \omega) = \frac{1}{2 i \pi} \int_{-\infty}^{+\infty}
   d\omega\: {\rm Tr}\,\bigl[G_R(\vec k , \omega)
  \bigr] \;,
}  
where we employed the definition of the spectral density function given in \eqref{spec_dens}.
In particular, we can deform the integration contour from the real axis to the infinite semi-circle in the upper half plane, i.e., we write $\omega = R\, e^{i\theta}$ with $0\leq \theta \leq \pi$ and send $R\to\infty$. 
Let us consider the second expression for the matrix $\xi(r,p)$ in (\ref{xibar}). The value of $\ov{\xi}_2$ on this semi-circle can easily be determined by noting that the limit $\omega\to\infty$ with $k$ finite is the same as $k\to 0$ and $\omega$ finite. Using (\ref{anaw}) this gives $\lim_{\omega\to\infty} \ov{\xi}_2 = const.$ on the infinite semi-circle. Let us denote this semi-circle as $C$. Then, we find 
\be\lab{sumrulez} 
\int_{-\infty}^{\infty} d\omega\, \rho(\vec k , \omega) = \frac{1}{2 i \pi} \int_C d\omega\: {\rm Tr}\,\bigl[G_R(\vec k , \omega)
\bigr]  = 1 \;,
\ee
iff $M<z/2$, regardless the value of $k$ (as long as it is finite). In this case, also the pole at $\omega=0$ does not contribute. Therefore we again find that the sum-rule is obeyed for $-z/2< M < z/2$. This generalizes the result in the previous section for $z=1$.  

In this general case we simply {\em assumed} that the Kramers-Kronig relation holds and then showed that the sum-rule is obeyed in the allowed range of $M$. 
For the particular case of $k=0$, however, we can also the Kramers-Kronig analytically. This is a straightforward generalization of discussion below equation (\ref{GRAdS1}) where for generic $z$, in place of (\ref{GRAdS1}) we have 
\be\lab{Gz}
G_{R}( \vec 0,\omega) = -\frac{1}{\omega -  \tilde{g}_{M,z}\:  \omega^{\frac{2M}{z}} e^{-i \pi (\frac{M}{z}+\half)}}\;.
\ee
The constant (\ref{gtilde}) in the present case generalizes as
\be\lab{gtildez}
\tilde{g}_{M,z} =   \frac{g_f}{Z} (2z)^{-\frac{2M}{z}} \frac{\Gamma\le(\half-\frac{M}{z}\ri)}{\Gamma\le(\half+\frac{M}{z}\ri)} r_0^{-2M+z+1} \;,
\ee
which is positive definite in the range $-z/2< M < z/2$ for $g_f/Z$ being positive definite. We again find a pole at $\omega=0$, a branch-cut on the lower imaginary axis and there is no pole in the upper half-plane of the principal sheet provided that 
\be\lab{Polez}
\tilde g_{M,z} >  0
\; .
\ee
Clearly, all non-analytic behavior is in the lower half-plane and the Kramers-Kronig relations are satisfied. For negative mass values, the self-energy is infrared divergent, which signals an IR singularity that 
may have interesting applications for condensed matter physics.  
 
One can also ask what happens beyond  the range $|M|<z/2$. Explicit calculations \cite{upcoming}  show that for $M>z/2$ the requirement of analyticity in the upper half plane can only be satisfied when $g_f/Z<0$. However the sum-rule integral (\ref{sumrulez}) then yields $-1$, which signals a violation of unitarity and which  is indeed consistent with the ``wrong" choice for the sign of $g_f/Z$. 
On the other hand, in the opposite range $M<-z/2$ one finds that there always is a singularity in the upper half plane, as a result of which both unitarity  and causality is violated. We leave the details of this calculation to \cite{upcoming}. 
 
 
 \section{Discussion}

In this work we introduced a method to compute correlation functions of elementary fermion fields that satisfy canonical commutation relations, within a strongly interacting CFT with an arbitrary dynamical scaling exponent $z$, in the context of the holographic correspondence.  
Our emphasis is to satisfy the sum-rules that are obeyed by these elementary fermion fields, as observed in 
 ARPES experiments. We determined the condition for satisfying the ARPES sum-rules 
imposed on the mass of the dual bulk fermion field as $-z/2 < M <z/2$ for an arbitrary dynamical exponent $z$. In the field theory this requirement corresponds to coupling the elementary field to an operator $\cO_-$ with scaling dimension above the unitarity bound and {\em relevant} in the IR. 

Let us expand a little more on the latter condition. It is clear from (\ref{Gz}) that demanding 
the interaction term be relevant in the IR is $M<z/2$. One can also check this directly in the corresponding field theory action: Such an interaction is given by $ \int dt\, d^{d-1}x\,  \tilde{g}\, 
\ol{\Psi}_+ \cO_-$ in the action. The question is whether the coupling $\tilde{g}$ is irrelevant or not in the UV. The scale transformation is $\vec{x} \to \Lambda^{-1} \vec{x}$, and $t \to \Lambda^{-z} t $. The weights of the elementary fermion $\Psi_+$, the operator $\cO_-$, and the volume term under this transformation are $(d-1)/2$, $\Delta_+$ and $-(d-1+z )$, respectively.   
Then the {\em dimensionless coupling} that one constructs from $\tilde{g}$ is $g = \tilde{g} \Lambda^{\Delta_+ - z - \frac{d-1}{2}}$.   For the beta-function of this coupling to be negative, i.e. being relevant in the IR, one should have $\Delta_+ <   z + \frac{d-1}{2}$, which indeed corresponds to $M<z/2$ using (\ref{scadim0}).

As a by-product of our analysis, we present 
analytic expressions for the fermion correlation functions in strongly interacting non-relativistic CFTs.\footnote{The special case of $z=2$, $M=0$ was studied 
in detail in \cite{Korovin}.}

There are various directions one can extend our work. The most immediate generalization involves turning on nonzero temperature and chemical potential \cite{upcoming}. The zeroth sum rule (\ref{contour-G})  should also be satisfied in the case of  nonzero temperature. This will be a non-trivial test of the generalization of our method to arbitrary temperatures. This step will be crucial of one desires to compare the results of holography with real ARPES data for systems with similar properties such as  bi-layer graphene.

Another question involves coupling the elementary field to more than one CFT operator. Indeed, this should be a more general situation in the ARPES experiments: the excited elementary fermion may interact with the strongly coupled CFT through many operators. This case was considered in \cite{FP} in a semi-holographic fashion. In order to study this situation in a more ``holographic" manner, one can imagine  turning on a source $\psi_+^\Delta$ for each CFT operator $\cO_-^\Delta$. In addition, if one turns on another elementary fermion, say $\chi_-$ on the UV cut-off surface, couple it to all of the sources on the boundary, and calculate the two-point function of $\chi_-$ instead of $\psi_+$, one may be able to achieve this goal.


\section*{Acknowledgments}
We thank Francesco Benini, Sean Hartnoll, Christopher Herzog, Sung-Sik Lee, John McGreevy,
David Tong, David Vegh and especially Koenraad Schalm for interesting discussions. This work was partly supported by the Netherlands Organization for Scientific Research (NWO) under the VICI grant 680-47-603.


\end{document}